\documentclass{article}

\usepackage{arxiv}

\usepackage[utf8]{inputenc} 
\usepackage[T1]{fontenc}    
\usepackage{hyperref}       
\usepackage{url}            
\usepackage{booktabs}       
\usepackage{amsfonts}       
\usepackage{nicefrac}       
\usepackage{microtype}      
\usepackage{graphicx}
\usepackage{doi}
\usepackage{blindtext}
\usepackage{enumitem}
\usepackage{epsfig}
\usepackage{graphicx}
\usepackage{amsmath}
\usepackage{amssymb}
\usepackage{multicol}
\usepackage{float}
\usepackage{mcite}
\usepackage[euler]{textgreek}
\usepackage{subcaption}
\usepackage{caption}
\usepackage{dblfloatfix}
\usepackage{cleveref}

\usepackage{physics}
\usepackage{siunitx}
\usepackage{cite}
\usepackage{braket}

\DeclareSIUnit\molar{\mole\per\cubic\deci\metre}
\DeclareSIUnit\Molar{\textsc{m}}

\setitemize[0]{leftmargin=*}
\setenumerate[0]{leftmargin=*}
\DeclareCaptionLabelFormat{cont}{#1~#2\alph{ContinuedFloat}}
\renewcommand{\headeright}{}
\renewcommand{\undertitle}{}
\renewcommand{\shorttitle}{Measuring the measurement problem}

\hypersetup{
pdftitle={},
pdfsubject={},
pdfauthor={},
pdfkeywords={},
}

\begin{document}
\title{Measuring the measurement problem: controlling decoherence with measurement duration in molecular MCB junctions}

\author{C.J. Muller
\\ \\
This research has, in its entirety, been privately conducted and funded by the author.
}
\maketitle

\begin{abstract}
We investigate the influence of the measurement duration on quantum coherence in molecular mechanically controlled break junctions operating in a tetrahydrofuran (THF) partially wet phase. These systems represent a distinct class of enclosed open quantum systems with unusually long decoherence times at ambient conditions, on the order of 1-20 \si{\milli\second}. By tuning the integration time of the current measurement in current-voltage (IV) characteristics, relative to the decoherence time, we observe a transition from quantum interference patterns, manifested as structured bands of data points, to classical behavior characterized by a single averaged response. This demonstrates that the duration of a measurement acts as a controllable parameter for probing quantum behavior in molecular junctions, offering new insights into decoherence dynamics in quantum mechanics.
\end{abstract}

\begin{multicols}{2}

\section{Introduction}
The measurement problem \cite{Karlsson2025} addresses the apparent discontinuity between the deterministic unitary evolution of a quantum state and the probabilistic outcome observed upon a measurement. Making a measurement or observation within the context of the Copenhagen interpretation implies that the quantum world is completely separated from our classical world. Here the classical world also includes the measurement equipment. A measurement is treated as an external, classical process, leaving the mechanism of wavefunction collapse unresolved. The measurement itself is a physical action, superimposed on the system it measures which implies that we can describe the two systems combined as a new system. Providing the measurement point in the equipment has not been observed by a human, we can add the human to the combined system again arriving at a higher-level new system. In this way we will end up with a Schr{\"o}dinger’s cat analogy. Decoherence theory provides for a partial resolution by describing how entanglement with an environment suppresses quantum interference, effectively yielding classical outcomes without invoking an explicit collapse.
\\ \\
Open quantum systems, where a quantum system interacts with its environment, are central to understanding decoherence \cite{Breuer2006,Modi2018}. Before any interaction between the quantum system and the environment has taken place the total wave function of the quantum system and environment combined can be described as:

\begin{equation}\label{eq.1}
    \ket{\psi}_{\text{OQS}}=\ket{\psi}_{\text{QS}}\ket{\psi}_{\text{E}}
\end{equation}

with $\ket{\psi}_{\text{OQS}}$ the wavefunction of the open quantum system, comprising the quantum system $\ket{\psi}_{\text{QS}}$ and the environment $\ket{\psi}_\text{E}$. When the quantum system starts to interact with the environment the two wavefunctions $\ket{\psi}_{\text{QS}}$ and $\ket{\psi}_{\text{E}}$ start to become entangled and they can no longer be separated, hence the appearance of equation \ref{eq.1} becomes much more complex. Physics describes the decohering of the quantum system, how the quantum mechanical character of the quantum system is lost, or transferred to the classical world, due to the interaction with the environment. In this way providing a quantitative connotation to the collapse of a wavefunction at the quantum to classical transition. Decoherence theory implies that not all coherence is lost at once but over a typical decoherence time, $\tau_c$, as information flows from the quantum system to the environment. As early as 1970 Zeh stressed that realistic quantum systems are never closed, never remaining coherent forever \cite{Zeh1970}. It was demonstrated that a quantum systems obeying the Schr{\"o}dinger equation becomes intensely entangled with its environment. It is this evolving entanglement that will suppress the quantum character, such as interference and coherence.
\\ \\
An insightful experiment revealing decoherence is by making use of the double-slit experiment \cite{Arndt2005}. When particles, electrons, atoms, or even molecular fullerenes are blasted at a double-slit in vacuum they will generate an interference pattern on a further located screen. As such displaying the quantum, Broglie wave, nature of the particles as they traverse the double-slit, even when the experiment is performed with one particle at the time. When we add a detection possibility to observe which of the two slits the individual particle has passed through, the interference pattern vanishes, showing only two lines on the screen right behind the two slits. The quantum nature of the particles is destroyed by observing which slit they went through, providing for a classical result only. 
\\ \\
The interesting part comes when the vacuum is reduced by inserting some gas. In principle we can measure the direction and speed of gas molecules that collide with the particles that traverse the slit. We would have to go to great lengths, putting detectors around the entire setup, however it is possible. This will provide us with the possibility to extract a slit 1 or 2 probability for the traversing particle. This possibility alone, without performing the exercise, is sufficient to fade the interference by decohering the quantum nature of the particles. This is crucial, the quantum nature disappears because of evolving entanglement where in principle we have an increasing amount of data available to determine a slit 1 or 2 probability. Hence killing the quantum effect and arriving at the classical world, similarly as in the above case when a detection possibility was added to the experiment. Here we observe in an experiment that the outcome can be tuned from coherence to decoherence or equivalent, from quantum to classical. This experiment contains all the ingredients that can be captured in a theoretical framework \cite{Schlosshauer2019}, which serves as a basis for describing the decoherence process of an open quantum system. In addition, the above experiment reveals why coherence is so fragile and easily destroyed, decoherence times can be extremely short, much shorter as compared to the times required to reach thermal equilibrium.
\\ \\
Here we show results from a specific type of open quantum system, the molecular MCB junction in the tetrahydrofuran partially wet phase. A bridging molecule between two electrodes represents the quantum system, the partially wet phase represents a controlled environment, a Faraday cage encloses the entire system from the larger environment \mcite{Muller2021,*Muller24}. We will call such a system an 'enclosed open quantum system'. Most studies assume Markovian dynamics with unidirectional information flow. Here, we focus on enclosed open quantum systems, where the environment is controlled and isolated, enabling strong non-Markovian effects and a bi-directional exchange of information. It has been unclear to this point under which exact conditions the enclosed open quantum system shows coherence. Here we experimentally demonstrate that the critical quantity is the measurement time in relation to the decoherence time. 

\begin{figure}[H]
    \centering
    \includegraphics[width=.49\textwidth]{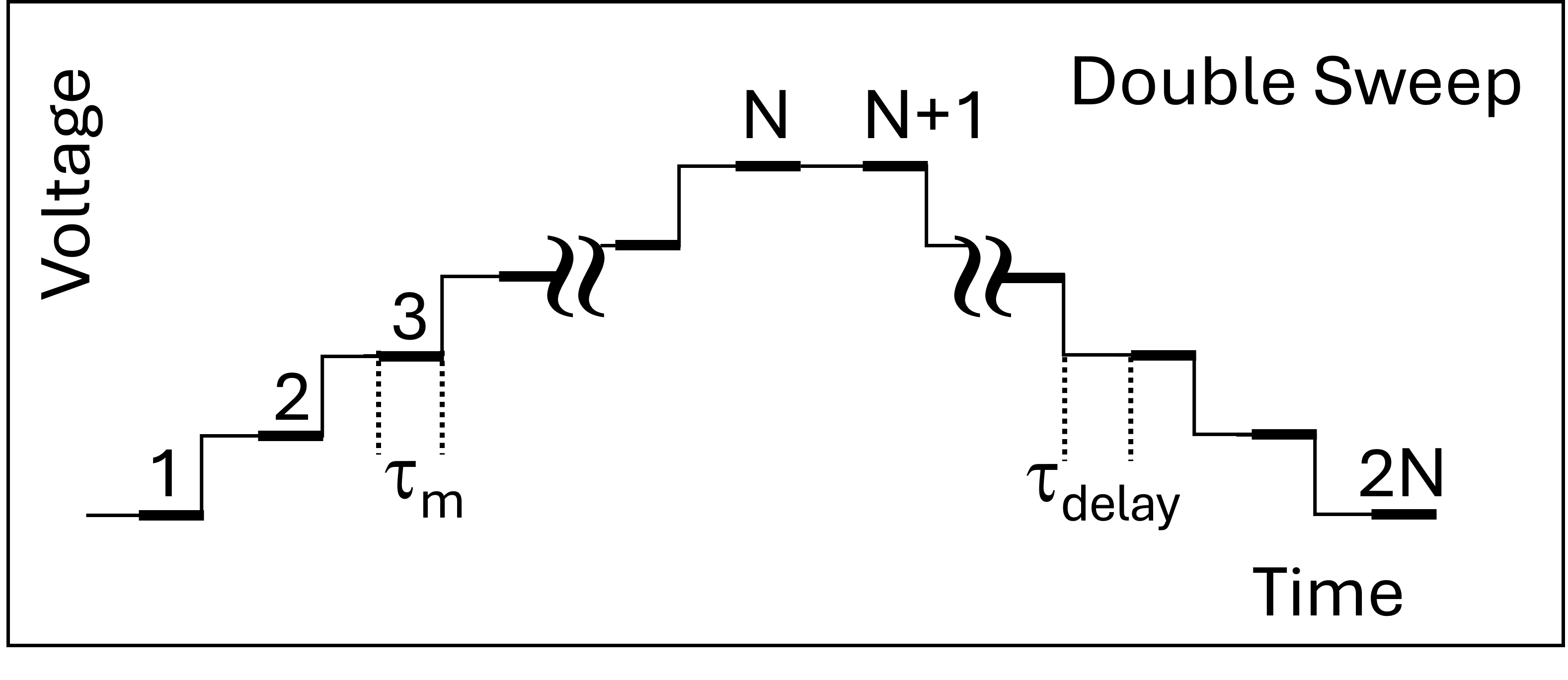}
    \caption{Schematic of a double scan consisting of 2N measurements, $\text{N}=1000$. A measurement consists of an integration time $\tau_m$ and a delay time $\tau_\text{delay}$. During $\tau_m$ the total charge that flows through the junction is integrated. During $\tau_\text{delay}$ the measurement current is calculated and stored, it also serves as some settling time.}
    \label{fig:fig1}
\end{figure}

\section{Experimental setup and results}
\begin{figure}[H]
    \centering
    \includegraphics[width=.47\textwidth]{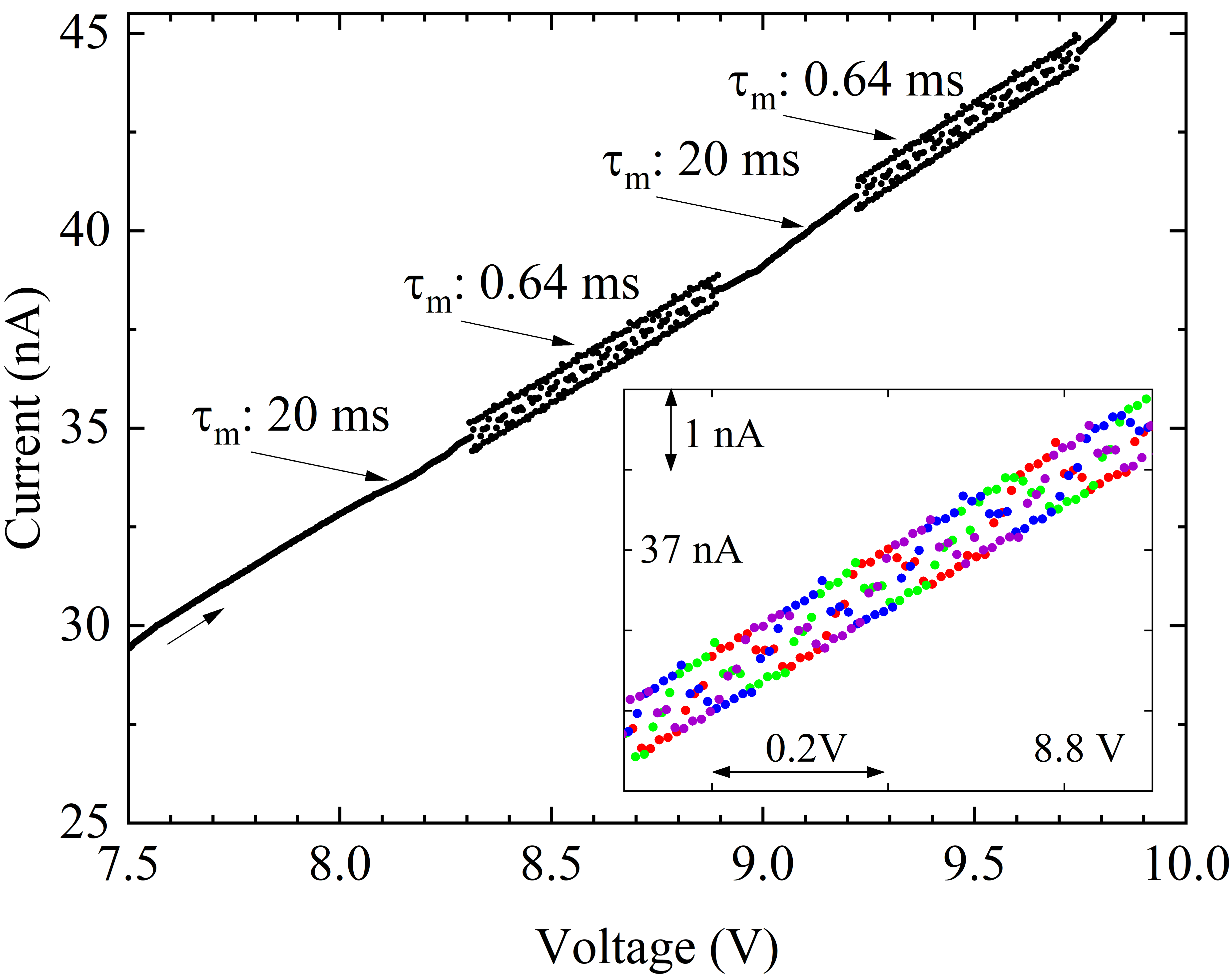}
    \caption{A characteristic IV curve from BBA1 where the measurement time is toggled between $\tau_m=\SI{20}{\milli\second}$ and $\tau_m=\SI{640}{\micro\second}$, indicating the sensitivity of quantum coherence in relation to the measurement time.}
    \label{fig:fig2}
\end{figure}

The experimental set up and operation has been described in detail in reference \mcite{Muller2021,*Muller24}, it allows for breaking a gold wire in a THF environment at ambient conditions. After a controlled drying process, a single molecule junction in the partially wet phase is created via self-assembly. The THF partially wet phase provides for a Faraday cage, offering the special conditions for enclosed open quantum systems. The measurements are performed on two bending beam assemblies (BBA), BBA1 in a benzene dithiol (BDT) THF solution and BBA2 in a THF only solution. Voltage biased IV characteristics consisting of 1000 data-points per scan are recorded in the partially wet phase. Typically, an IV curve contains two scans, one with an increasing voltage and one with a decreasing voltage. There is no delay at the point of scan reversal. 
\\ \\
A measurement of the MCB device is defined by the duration of the integration time $\tau_m$ as detailed in fig \ref{fig:fig1}. The current is calculated by integrating the charge that flows through the junction divided by $\tau_m$. The delay time $\tau_\text{delay}$ is used to calculate the current from the previous measurement, store this value, and for settling time of the new voltage value prior to the next measurement. A single scan ($\text{N}=1000$) lasts 45 seconds for $\tau_m=$ \SI{20}{\milli\second} and 5 seconds for $\tau_m=$ \SI{640}{\micro\second}. High speed measurements, $\tau_m=$ \SI{640}{\micro\second}, lead to a lower resolution and a higher measurement noise band, while slower measurements, $\tau_m=$ \SI{20}{\milli\second}, exhibit the opposite behavior. The noise bandwidth has been measured on a fixed \SI{50}{M\Omega} resistor to be \SI{15}{pA} for $\tau_m=$ 20 ms scans and \SI{50}{pA} for $\tau_m=$ \SI{640}{\micro\second} scans. Data presented here result from “fast” measurements with $\tau_m=$ \SI{640}{\micro \second} for all graphs except for fig \ref{fig:fig2} where $\tau_m$ is toggled between \SI{640}{\micro\second} and \SI{20}{\milli\second}.

\begin{figure}[H]
    \centering
    \includegraphics[width=.47\textwidth]{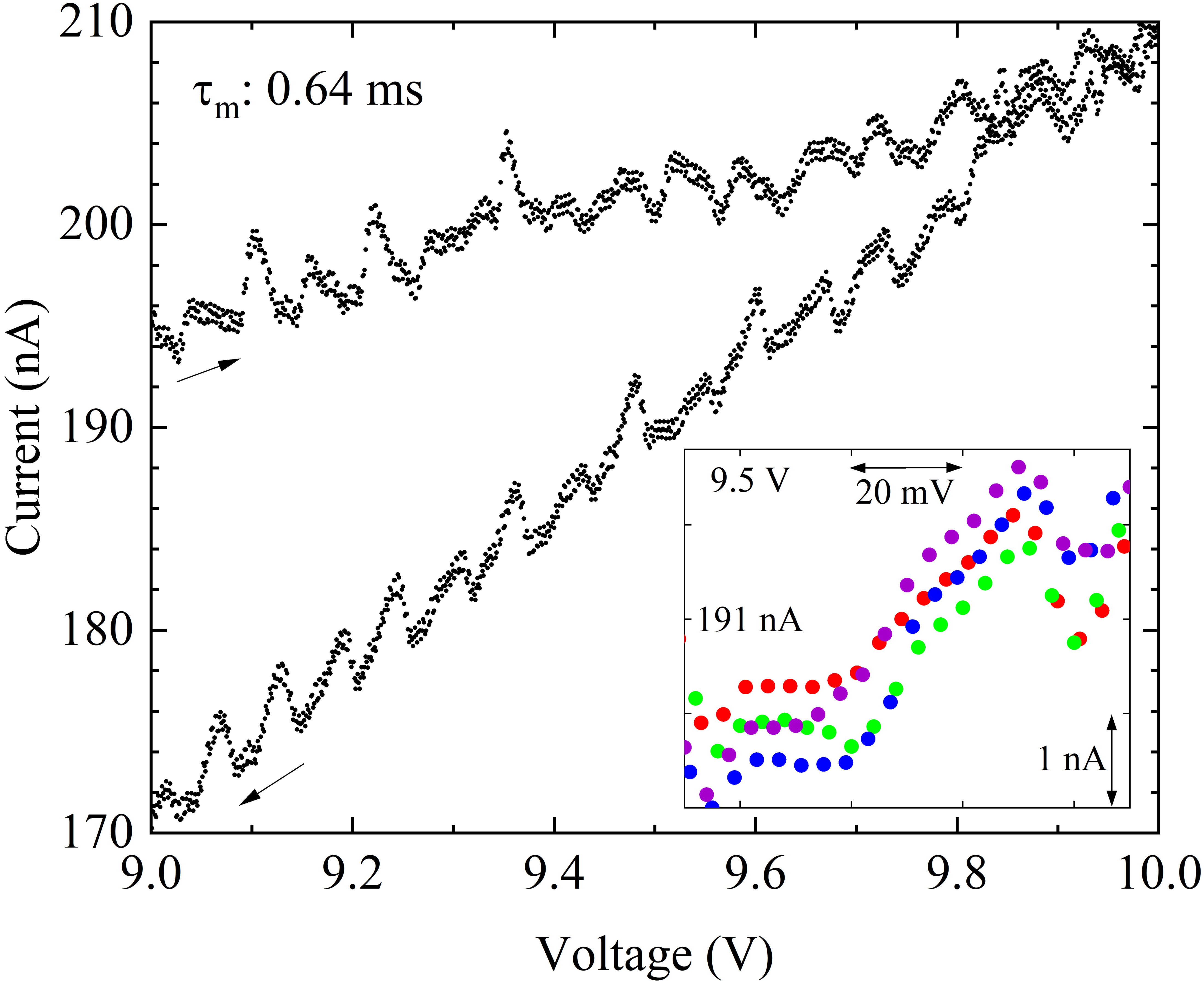}
    \caption{First IV curve from BBA2 showing a spiky baseline with a superimposed finite bandwidth. Inset: four interleaved groups, \SI{1}{\milli\volt} per step, same color separation of \SI{4}{\milli\volt}.}
    \label{fig:fig3}
\end{figure}

Fig. \ref{fig:fig2} shows the IV curve under conditions where $\tau_m$ has been toggled in BBA1. The IV curve shows clear structure for the $\tau_m=$ \SI{640}{\micro\second} sections where data points seem to favor positions at the extremes and the middle of a band. Every fourth data point creates the data lines at the extremes of the band, this holds for all main IV curves in fig. \ref{fig:fig2}, \ref{fig:fig3} and \ref{fig:fig5}. Because of this, the data has been separated into four groups consisting of group1: point 1, 5, 9... group2: point 2, 6, 10... and so forth. For the colored data in all shown graphs the measurement sequence in the scan direction is the same: green, blue, purple, red, green and so forth. 

\begin{figure}[H]
    \centering
    \includegraphics[width=.47\textwidth]{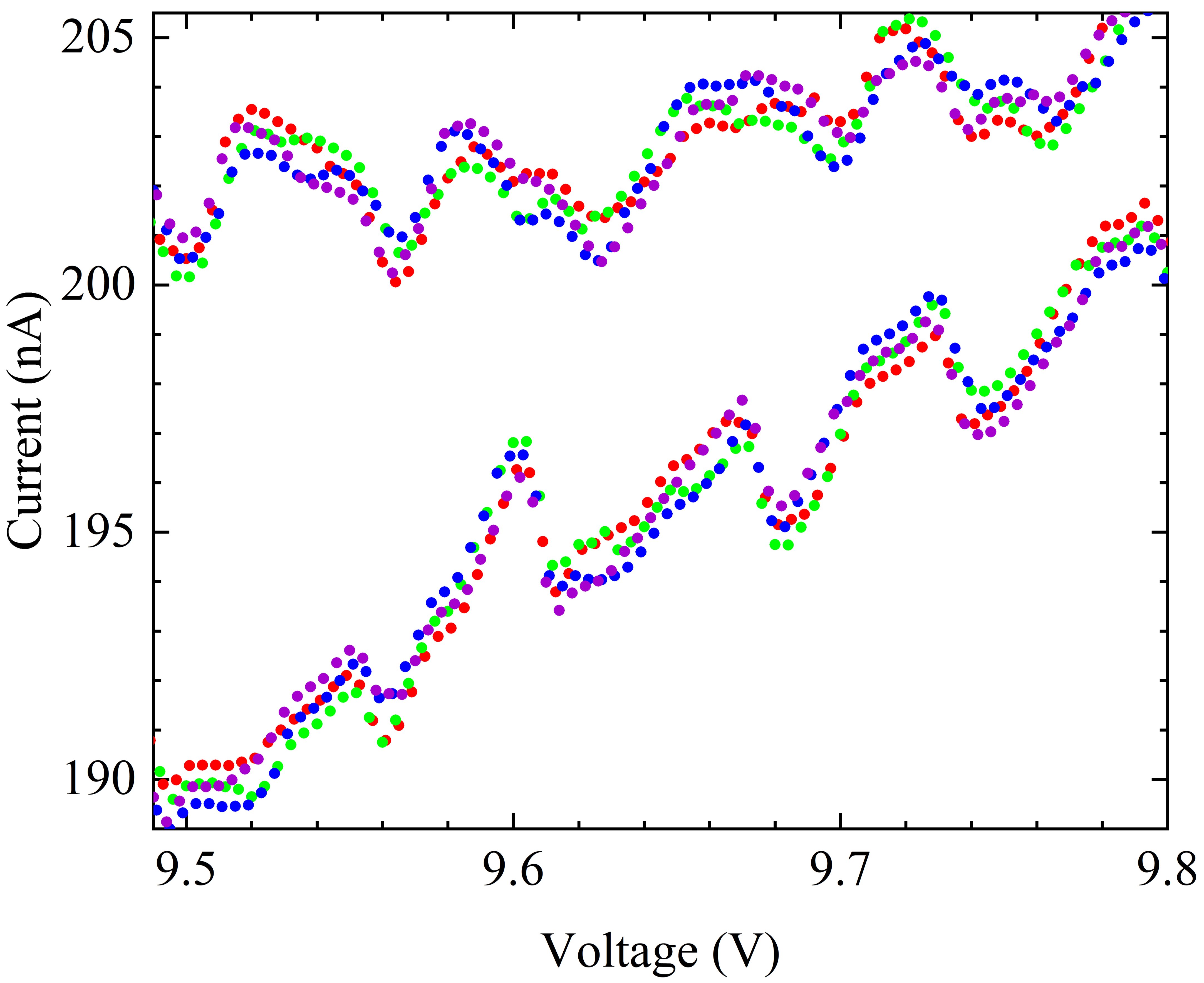}
    \caption{A section of fig. \ref{fig:fig3}, highlighting alternation of group colors between band edges.}
    \label{fig:fig4}
\end{figure}

\begin{figure}[H]
    \centering
    \includegraphics[width=.47\textwidth]{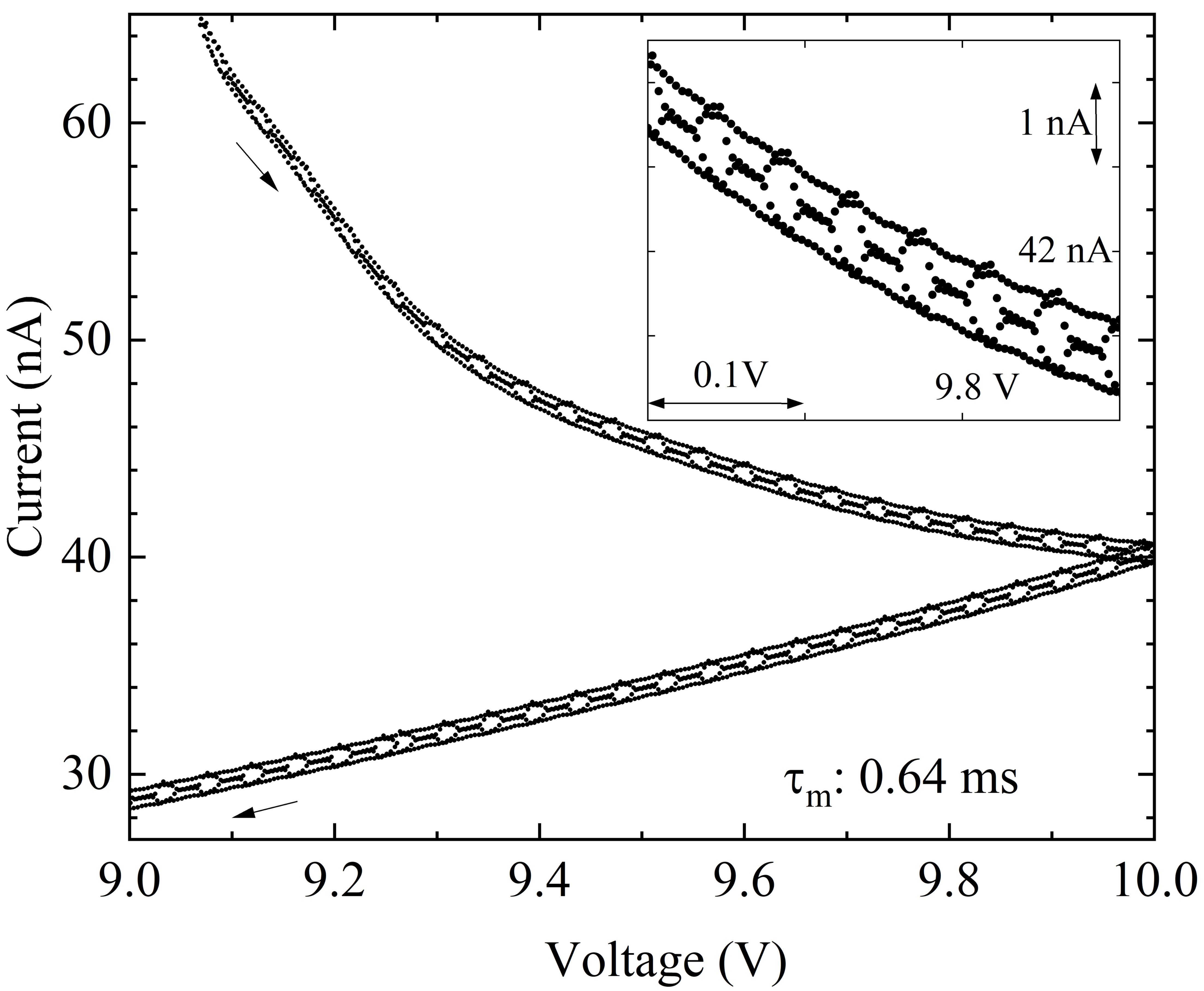}
    \caption{Second IV curve from BBA2 with a consistent bandwidth and clear repetitive structure, visible in both the main panel and inset.}
    \label{fig:fig5}
\end{figure}

The inset in fig. \ref{fig:fig2} shows a part of the data with a repetitive pattern, where colored data points change from one extreme to the opposite extreme via the position in between. The thousand points of this curve extend over \SI{2.5}{\volt}. The measurement equipment caters only for integer mV values, thus the voltage separation between sequential data points alternates, \SI{2}{\milli\volt}, \SI{3}{\milli\volt}, \SI{2}{\milli\volt}, \SI{3}{\milli\volt}, and so forth. This implies a voltage separation of \SI{10}{\milli\volt} between same color points. 

Fig. \ref{fig:fig3} recorded for BBA2 shows data that follow a spiky curve, with a superposition of a band of data points. In this instance the separation between sequential data points is 1 mV thus the difference between same color points is 4 mV. In the inset of fig. \ref{fig:fig3}, the data are once more grouped in four sets as described before, revealing the repetitive nature of the data. Fig. \ref{fig:fig4} shows another example of the systematic behavior of the data band, superimposed on the spiky curve. Colored groups of data points alternate between the two current extremes and the middle position of the data band. The band of data points typically has an amplitude of \SI{1}{\nano\ampere}, in fig. 2, 3 and 5. The structure of the colored groups has a periodicity of 160, 150, and 172 mV which is equivalent to \SI{0.32}{s}, \SI{0.75}{s}, and \SI{0.86}{s} at the fast scan-speed of \SI{500}{\milli\volt/\second} for fig. \ref{fig:fig2} and \SI{200}{\milli\volt/\second} for fig. \ref{fig:fig3} and \ref{fig:fig4}.

\begin{figure}[H]
    \centering
    \includegraphics[width=.47\textwidth]{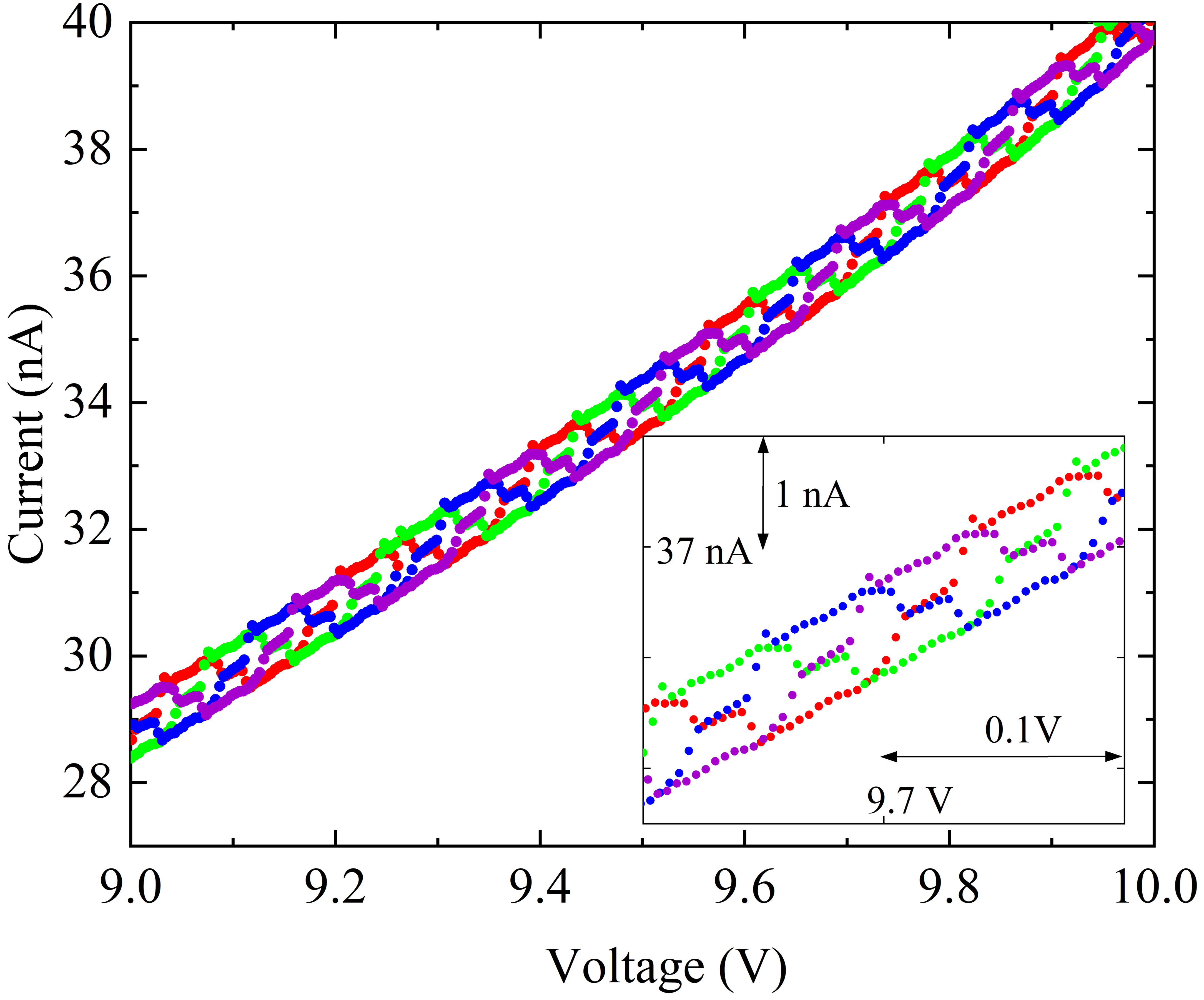}
   \caption{The lower branch of fig. \ref{fig:fig5} split into four groups reveals a recurring pattern.}
    \label{fig:fig6}
\end{figure}

\begin{figure}[H]
    \centering
    \includegraphics[width=.47\textwidth]{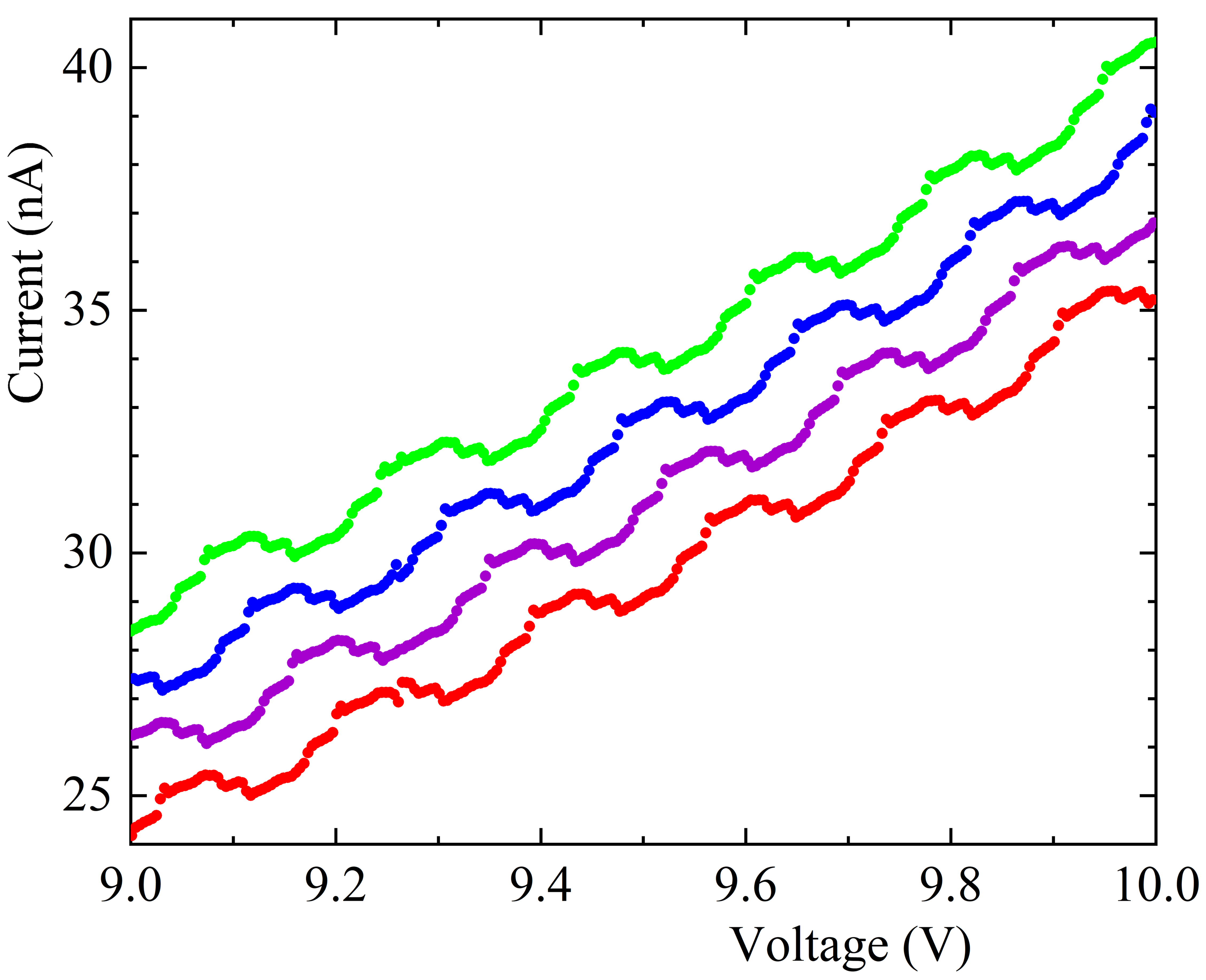}
    \caption{A further breakdown of fig. \ref{fig:fig6}: blue, purple, and red groups offset by \SI{-1.5}{\nano\ampere}, \SI{-3}{\nano\ampere}, and \SI{-4.5}{\nano\ampere} respectively show phase-shifted similarity.}
    \label{fig:fig7}
\end{figure}

\begin{figure}[H]
    \centering
    \includegraphics[width=.47\textwidth]{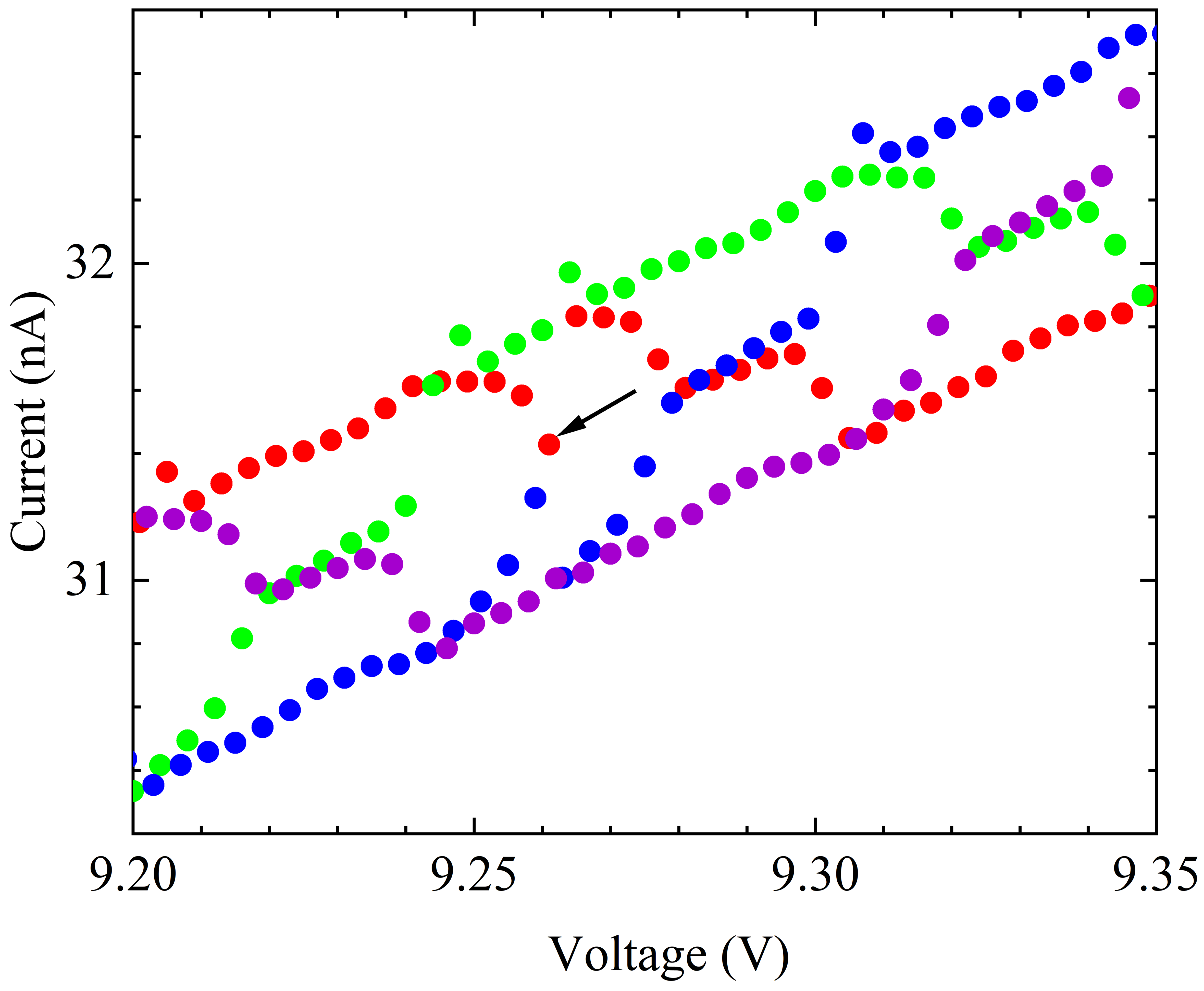}
    \caption{Enlargement of the pattern breakdown in the lower curve from fig. \ref{fig:fig5} near \SI{9.25}{\volt}. The shifted pattern starts at the off-pattern red point indicated by the arrow, which normally would have been located at the upper current extreme. After this the shifted pattern continues as before towards lower voltages.}
    \label{fig:fig8}
\end{figure}

Another IV curve from BBA2, fig. \ref{fig:fig5}, shows a typical IV curve towards the end of the partial wet phase with a relative fast reduction in conductivity. Also here a band of data points is visible in the entire curve. A section has been enlarged in the inset. Similarly, as before 4 groups of data were created which is shown for the lower curve only in fig. \ref{fig:fig6}. The inset in fig. \ref{fig:fig6} shows an enlargement of a section. This graph shows an equivalent repeating pattern of data points as well. This becomes even clearer in fig. \ref{fig:fig7} where each group is separately indicated, with the blue purple and red groups offset by \SI{-1.5}{\nano\ampere}, \SI{-3}{\nano\ampere}, and \SI{-4.5}{\nano\ampere} respectively. The data sets of the various groups look identical but shifted in time. In the main graph from fig. \ref{fig:fig5} a distinct black and white pattern repeats itself over the entire curve. In the lower curve, however close at \SI{9.25}{V}, it shows that the pattern is broken. Two white spots are close together, this section is enlarged in fig. \ref{fig:fig8}.  At the broken pattern the voltage difference between the two repeated red/blue point sections spanning the two current extremes of the band, equals \SI{20}{mV} or equivalently \SI{100}{ms}. After the broking pattern, the scan towards lower voltages follows the regular pattern once again.

\section{Theoretical background and discussion}
The data in fig. \ref{fig:fig2} are exemplary for toggling $\tau_m$, there have been at least over a dozen observations of the same effect in as many different bending beam assemblies. Thus, this seems a generic effect in THF partially wet phase molecular junctions. A band of data points has been observed before \mcite{Muller2021,*Muller24}. However, this data was recorded with $\tau_m=$ \SI{20}{ms} and showed either two interleaved sinusoidal curves with varying periodicity and fixed amplitude or, three or four interleaved oscillating curves. At these long integration times this data occurred less frequently, we named this effect quantum effect 1, (qe1). Here we observe four interleaved groups of data with a fixed group sequence. In general, a faster measurement leads to a wider measurement noise band. That is, however, not the origin of the data displayed here. First, the data cannot be considered as noise, it provides a regular pattern within the bandwidth. Second, the expected \SI{50}{\pico\ampere} measurement noise for these measurements is well below the \SI{1}{\nano\ampere} bandwidth structure.
\\ \\
In fig. \ref{fig:fig3}, the overall shape of the curve resembles the repetitive spiky pattern in reference \mcite{Muller2021,*Muller2024}. This structure was attributed to a continuously increasing or decreasing force on the bridging molecule because of an increasing or decreasing partially wet phase layer which was called quantum effect 2, qe2. Both qe1 and qe2 have been described in more detail in the same reference. We attribute the pattern in fig \ref{fig:fig3} also to qe2. Superimposed on the spiky base curve a band of data points shows in fig. \ref{fig:fig3} and \ref{fig:fig4}. This data thus suggests that the two quantum effects, qe1 and qe2 can coexist in one IV curve.
\\ \\
The smooth IV curve in fig. \ref{fig:fig5} facilitates an extraction of the data pattern, shown in fig. \ref{fig:fig6} and \ref{fig:fig7}, which display the level of regularity and similarity of the different data groups. Grouping the data into four sets and after providing an offset as is displayed in fig \ref{fig:fig7} these curves look identical, only shifted in time! The shape of these curves deviates from the earlier discussed sinusoidal or oscillatory interleaved curves. In fig. \ref{fig:fig7} curves have a fixed periodicity and data-points seem to favor the extreme and middle positions of the band. This seems to be generic for fast speed measurements. The broken pattern in fig. \ref{fig:fig8} provides an opportunity to see how nature deals with small mistakes. The effect turns out to be a “glitch” with one point (see arrow), after which the regular pattern continues as before. Thus, this relates to a pattern shift. The pattern may be defined by the interference of the measurement frequency and the two-way information diffusion process between the molecule and the controlled environment, the time $\tau_c$. In this way surviving a glitch with a resulting pattern shift. 
\\ \\
A clear image emerges of the conditions under which the band of data-points appears. Reducing the measurement time favors the band of data-points to occur. For long measurement times the data band still shows occasionally, but far less frequent and in far less BBA data sets. Here we argue that with the measurement time we have a knob which we can tweak, either to reduce or increase $\tau_m$ below or above the decoherence time of the system. For $\tau_m<\tau_c$ the response of the quantum system shows a band of data points, for $\tau_m>\tau_c$ a line of data points results. This implies that the typical decoherence time of these quantum systems is in the \SI{1}{ms} to 20+ \si{ms} range at ambient conditions. This is based on many experimental observations with $\tau_m=$ \SI{640}{\micro\second} or $\tau_m=$ \SI{20}{ms}.
\\ \\
For enclosed open quantum systems due to observed strong non-Markovianity, next to the flow of information from the quantum system to the controlled environment, there must be a flow of information from the controlled environment to the quantum system. The flow of information between the quantum system and the controlled environment is treated as a diffusion process with a typical time constant $\tau_c$. The quantum system is in equilibrium with the controlled environment, performing a measurement acts on the controlled environment and will disturb this equilibrium. Let’s assume the quantum system consists of a molecule with states $\ket{A}$ and $\ket{B}$ in a superposition, $\ket{\psi}=c_A\ket{A}+c_B\ket{B}$ where $c_A$ and $c_B$ are the probability amplitudes: $c_A^2+c_B^2=1$. State $\ket{A}$ has leaked $\alpha\ket{A}$ to the controlled environment. A measurement will erase $\alpha\ket{A}$ from the environment, this will proliferate through the measurement system and finish as a data point in the measurement equipment. This process will distort the equilibrium in the controlled environment. Since $\ket{A}$ character is depleted $\ket{B}$ character will predominantly flow from the controlled environment to the quantum system in a typical time $\tau_c$ subsequently the $\ket{B}$ vector of the quantum system is bolstered and will start to leak to the controlled environment over the same typical time $\tau_c$. As long as $\tau_m<\tau_c$ we will measure the two distinct states. The time resolution of the measurement is below the decoherence time, enabling a measurement of a separate $\ket{A}$ and $\ket{B}$ state. In the other extreme where  $\tau_m>\tau_c$ at the start of the integration time $\alpha\ket{A}$ will be erased. At some point $\ket{B}$ character starts to flow to the quantum system and will leak $\beta\ket{B}$ to the controlled environment. Multiple cycles are measured during the same integration interval, hence measuring multiple $\alpha\ket{A}$ and $\beta\ket{B}$ associated current values. Thus, the measurement will display an average of the $\alpha\ket{A}$ and $\beta\ket{B}$ correlated current values. The separate quantum values get averaged and thus lost.
\\ \\
How should the data pattern be interpreted? The duration of a single fast measurement is comparable to $\tau_c$. The interplay between the measurement equipment and the enclosed quantum system may lead to interference between the data repetition rate and the information diffusion process. To increase our knowledge of the pattern we will need more variation in the measurement-frequency to study the resulting response from the enclosed quantum system. At this stage the exact nature of the pattern remains unexplained. 
\\ \\
What makes enclosed open quantum systems special? Why is the decoherence time exceptionally long? We are only at the beginning of a very interesting phase. We may have underestimated the influence of Heisenberg or similarly the environment on the behavior of devices up to this point. 

\section{Conclusions}
The molecular MCB junction in the partially wet phase provides for a special type of open quantum system, the enclosed open quantum system. In this type of device information flows in both directions between the quantum system and the controlled environment. Typical decoherence times are in the \SI{1}{ms} to 20+ \si{ms} range at ambient conditions. With the measurement time we have a powerful knob, enabling adjustment in relation to the decoherence time of the system and in this way, we can study the transition from coherence to decoherence. It has been shown that the quantum effect related to a continuous increasing or decreasing strain on the bridging molecule can coexist with the quantum effect related to the band of data points resulting from the response of the enclosed open quantum system to repetitive measurements.


\bibliographystyle{unsrt}
\bibliography{references}

\end{multicols}

\end{document}